\documentclass[acmtog]{acmart}

\usepackage{xcolor}
\usepackage{colortbl}
\usepackage{bbm}
\usepackage{algorithm}
\usepackage{algpseudocode}
\usepackage{wrapfig}
\usepackage{blkarray}

\DeclareMathAlphabet{\mathpzc}{OT1}{pzc}{m}{it}

\newcommand{\ty}[1]{}
\newcommand{\Honglin}[1]{}
\newcommand{\nicetohave}[1]{}

\newcommand{\real}{{\rm I\!R}}

\newcommand{\vc}[1]{\mathbf{#1}}
\newcommand{\mat}[1]{#1}

\AtBeginDocument{%
  \providecommand\BibTeX{{%
    \normalfont B\kern-0.5em{\scshape i\kern-0.25em b}\kern-0.8em\TeX}}}

\setcitestyle{square}

\setcopyright{none}
\begin{document}

%\title{Mixed Variational Finite Elements for Implicit, Position-Based Simulation of Solid Bodies}
%\title{Mixed Variational Finite Elements for Implicit, Position-Based Simulation of Solid Bodies}
%\title{Mixed Variational Finite Elements for Implicit, General-Purpose Simulation of Solid Bodies}
\title{Mixed Variational Finite Elements for Implicit, General-Purpose Simulation of Deformables}

\author{Ty Trusty}
\affiliation{%
  \institution{University of Toronto}
  \country{Canada}
}

\author{Danny M. Kaufman}
\affiliation{%
  \institution{Adobe Research}
  \country{USA}
}

\author{David I.W Levin}
%\authornotemark[3]
\affiliation{%
  \institution{University of Toronto}
  \country{Canada}
}

\renewcommand{\shortauthors}{Levin}

\begin{abstract}
    We propose and explore a new, general-purpose method for the implicit time integration of elastica.
Key to our approach is the use of a mixed variational principle. In turn its finite element discretization leads to an efficient alternating projections solver with a superset of the desirable properties of many previous fast solution strategies.
This framework fits a range of elastic constitutive models and remains stable across a wide span of timestep sizes, material parameters (including problems that are quasi-static and approximately rigid). It is efficient to evaluate and easily applicable to volume, surface, and rods models. We demonstrate the efficacy of our approach on a number of simulated examples across all three codomains.
\end{abstract}




\begin{teaserfigure}
  \includegraphics[width=.95\textwidth]{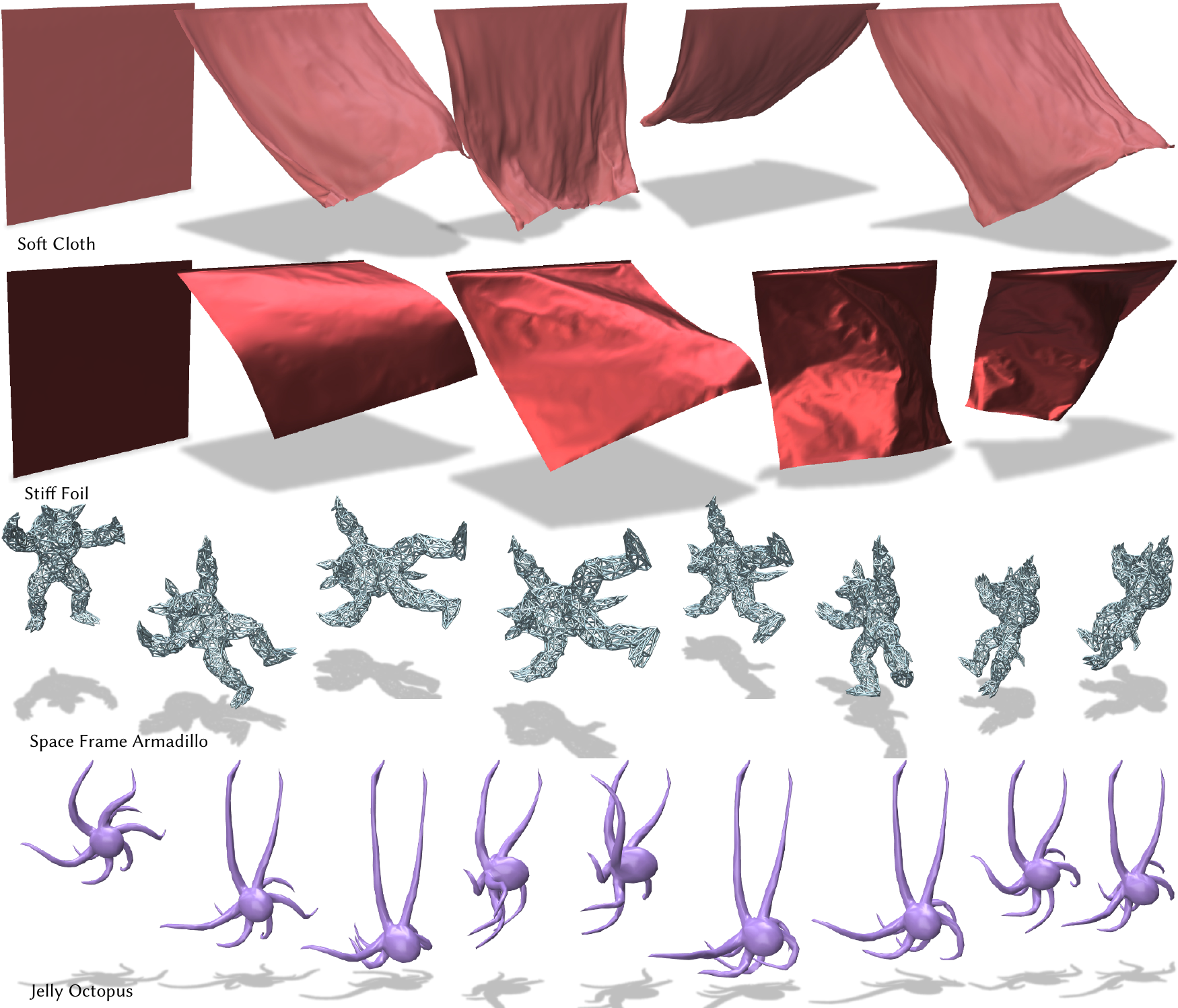}
  \caption{Interactive dynamic simulations of elastic and rigid solid objects (volumes, shells and rods) using our unified mixed variational solver.}
  \label{fig:teaser}
\end{teaserfigure}

\maketitle

\section{Introduction}
In this paper we explore the use of a mixed variational principle to build an efficient and general-purpose simulation algorithm for the physics-based animation of elastica.

Standard approaches for the implicit time integration of continua discretize with finite differences in time and finite elements in space. Recent methods often leverage the observation that, for these implicit time integration choices, each individual time step solve can then be cast as a minimization problem.
In turn, the applied strategy for solving these optimization problems then leads to a wide range of well-known simulation algorithms\ \cite{Li:2019:DOT}. 
For example, a ``standard'' finite element approach involves minimizing an implicit integration energy via Newton's method while solving the bottleneck of inner linear-systems solves either via direct or iterative methods.
Extended Position-Based Dynamics replaces standard direct or iterative solvers with iterations (e.g., GS, Jacobi, and/or SOR) acting on the dual variables (constraint forces) while Projective Dynamics and its more recent generalizations apply various forms of ADMM-type solvers to split, augmented Lagrangian forms.

Despite their common variational origin, implicit solvers for elastica exhibit a wide range of features and limitations, and so tradeoffs. 
Standard Newton-type approaches and ADMM-based methods (including Projective Dynamics) exhibit various difficulties simulating stiff materials. 
On the other hand, Position-Based Dynamics, while fast and stable (even at high stiffness), lacks a direct correspondence to a discretization model and underlying PDE -- it is then often unclear how to convert general material moduli and models to fit its constraint-based formulation. 

Arguably an implicit time integrator for physics-based animation should 
\begin{itemize}
\item support general elastic constitutive models;
\item easily adapt to volumetric, surface and rod simulation;
\item offer stability for a wide range of materials including those stiff enough to be effectively rigid;
\item remain stable for large, frame-rate sized time steps; and
\item provide efficient solutions of every time step.
\end{itemize}
Existing popular methods, as covered briefly here and below, are often custom-specialized to a small subset of these properties.
Here, we explore a mixed variational finite-element model and an efficient solver for it, that covers the full spread of the above target properties.

\section{Related Work}

Implicit time integrators, especially backwards Euler, are ubiquitous tools in simulating elastica. %computer graphics. 
%They are direcly introducing a velocity variable into Newton's second law, resulting in a nonlinear system of coupled first-order differential equations.
In computer graphics implicit integration steps are often solve via 
%This system of equations was solved via 
Newton's Method~\citep{10.1145/176579.176580} or else via a single-iteration, linearly implicit approximation~\citep{10.1145/280814.280821}.
A range of applied forces can, in turn, be derived variationally from distortion metrics~\citep{10.1145/176579.176580} and mechanical conditions on the simulated objects~\citep{10.1145/280814.280821}.
More recent approaches often derive and implement Backward Euler time integration from a variational perspective~\citep{10.1145/1964921.1964967,hairer2014energy} which, once fully discretized in time (via finite differences), and space (via finite elements), yields a nonlinear minimization to be solved to update each forward time step in a simulation.
Efficient, robust, and accurate solutions of this optimization problem now lies at the heart of recent, physics-based animation research.

A default solution approach is to apply Newton's method with a direct solver to handle the resulting per-iteration sequence of linear systems solves. In turn iterative linear solvers~\citep{SNHkim} often offer better linear solve performance and memory usage with the trade-off of overall slow convergence for stiffer material models. 
A exceedingly efficient alternative approach is to apply primal-dual solver strategies which model elastic forces as constraints. Resulting dual optimization problems are then (approximately) solved to compute each forward time step. An extremely successful and effective application of this strategy is the Position Based Dynamics (PBD) algorithm~\citep{MULLER2007109} which acts on compliant constraints (similar to the condition energies from \citet{10.1145/280814.280821}), with fast, local iterative updates to generate approximating solutions for the dynmaics.
Extended Position Based Dynamics (XPBD) extends PBD with a quadratic compliant formulation~\citep{servin2006interactive} to establish a relationship between PBD constraints and some material models~\citep{10.1145/2994258.2994272} and enables the simulation of both rigid and quasi-rigid bodies~\citep{10.1111:cgf.14105}.
However the local nature of the PBD updates, and its dependence on iteration tuning often necessitates additional complexity in achieving good convergence in high resolution meshes and large domains~\citep{10.1145/3099564.3099574}. 
XPBD type solves also arise as intermediate steps in non-smooth Newton's Methods for contact and friction~\citep{DBLP:journals/corr/abs-1907-04587}.

Projective Dynamics (PD) ~\citep{10.1145/2601097.2601116} applies an ADMM-type algorithm~\citep{overby2017admmpd} to minimize the integration energy of a subset of deformation energies. This strategy has since been fully generalized by Overby et al.\ \citep{overby2017admmpd} to a full ADMM model covering a complete range of hyperelastic energies.
These methods incorporate a highly effective global projection step which helps balance locally solved forces across an entire meshed domain and lead to incredibly stable behavior when simulating nonlinear elastica. 
However, this comes with a trade-off. When simulating nonlinear materials these ADMM strategies suffer from slow and even nonconvergence~\cite{Li:2019:DOT}, algorithmic parameter tuning questions, and challenges when it comes to simulating stiffer materials.
Most recently these parameter tuning and convergence questions have been better resolved with the trade-off of more expensive global and local update steps~\citep{brown2021wrapd}.

While recent work, for which the above is just a small sampling, has rapidly explored improvements and  tradeoffs at the solver level (and occasionally in choice of time-discretization) the underlying variational energies being optimization have  largely remained the same. 
Closer to our exploration mixed variational principles have previously been explored for deriving new implicit \emph{time} integrators. The Hamilton-Pontryagin principle~\citep{10.5555/1218064.1218071} has been applied to resolve position and displacement as independent discrete time-stepped variables coupled via constraints, in order to derive discrete variational time integrators. 
Here we are instead motivated by Reissner's \shortcite{10.1007/bf01177113} approach, which decouples deformation and displacements.
Embarking from a Reissner-motivated mixed variational principle, we apply design choices motivated from prior work in both computer graphics and numerical optimization. This naturally leads us to an efficient elastodynamics solver that, while superficially resembling fast solvers in geometry processing~\citep{Jacobson-12-FAST}, exhibits a superset of features currently available in physics-based methods. 
These include compatibility with general elastic constitutive models, stability for a wide range of time step sizes and material parameters (including problems that are quasi-static and approximately rigid), as well direct applicability to volume, surface and rod simulation. We demonstrate the efficacy of our method on a number of examples simulated at interactive rates.

\section{Methods}
\fbox{\begin{minipage}{\columnwidth}
\textbf{Note: } We use tensor notation in our derivations. 
Basic tensors and operations are described in our supplemental material. 
It's highly recommended to have it handy upon first reading.
\end{minipage}}
\vspace{1em}

Given an input finite element discretization of a domain with $|T|$ elements and $|V|$ vertices, the standard, position-level optimization form of backwards Euler time integration~\citep{hairer2014energy} is
\begin{equation}
    \vc{q}^{t+1} = \arg\min_{\vc{q}} \frac{1}{2h^2}\vc{a}\left(\vc{q}\right)^T\mat{M}\vc{a}\left(\vc{q}\right) + U\left(\vc{q}\right),
    \label{eq:implicit}
\end{equation}

where $\mat{M}\in\real^{3|V|\times3|V|}$ is a mass matrix, $\vc{a}\left(\vc{q}\right) = \vc{q} - 2\vc{q}^t + \vc{q}^{t-1}$, all $\vc{q}^k\in\real^{3|V|}$ are stacked vectors of deformed vertex positions with $t$ and $t-1$ denoting the current and previous time steps, and $U$ is an elastic potential energy constructed by integration of a hyperelastic strain energy density, $\psi$, over the domain.

We immediately depart from standard derivations by taking a step back to the spatially continuous setting. We swap $U$ with a mixed variational potential~\citep{10.1007/bf01177113}, custom-constructed to explicitly account for local rotation coordinates,
\begin{equation}
    \tilde{U} = \int_\Omega \psi\left(\mat{S}\left(\vc{X}\right)\right) - \mat{\lambda}\left(\vc{X}\right):\left(\mat{R}\left(\vc{X}\right)\mat{S}\left(\vc{X}\right) - \mat{F}\left(\vc{X}\right)\right)d\Omega.
    \label{eq:mixed_continuous}
\end{equation} Here $\Omega$ is the undeformed domain of the modeled object, $\mat{F}\in\real^{3\times3}$ is the deformation gradient at position $\vc{X}\in\real^3$ in the undeformed domain, $\mat{R}\in\mbox{SO}^3$ are local rotation matrices and $\mat{S}\in\real^{3\times3}$, are  \emph{symmetric} local deformations. Lagrange multipliers, $\mat{\lambda}\in\real^{3\times3}$ then enforce consistency between deformation gradients and $\mat{R} \mat{S}$.
Our inclusion of rotations, enables our resulting solver to explicitly track local rigid motions, improving convergence, especially in stiff material simulations~\citep{brown2021wrapd}.
Standard potential energies for body forces can be added to this potential, though we omit them here for sake of brevity.

Returning to the discrete setting we construct a mixed potential on the simulation mesh by discretization (here, for clarity, using a tetrahedral mesh) with piecewise linear finite elements: 

\begin{equation}
    \begin{split}
        \tilde{U}^D = & \sum_{i=1}^{|T|}\left(\psi\left(\mat{\vc{C}}\cdot\bar{\mat{P}}_i\vc{s}\right)\right. \cdots \\
                      & - \left(\mat{\vc{B}}\cdot\mat{P}_i\vc{l}\right):\left(\mat{R}_i\left(\mat{\vc{C}}\cdot\bar{\mat{P}}_i\vc{s}\right) - \left.\left(\mat{\vc{D}}^4\cdot\mat{N}_i\vc{q}\right)G^3_i\right)\right)dv_i.
    \end{split}
    \label{eq:mixed_discrete}
\end{equation} Here $|T|$ gives number of tetrahedra and subscript $i$ indexes terms for the $i^{th}$ tetrahedron.  
Variables $\vc{q}\in\real^{3|T|}$, $\vc{s}\in\real^{6|T|}$ and $\vc{l}\in\real^{9|T|}$ are stacked vectors of deformed vertex positions, entries of $S$, and entries of $\lambda$ respectively, and $\mat{R}_i$ are local rotation matrices.
Next, $\bar{\mat{P}}_i$ (resp. $\mat{P}_i$ and $\mat{N}_i$) are sparse selection matrices that select from $\vc{s}$ (resp. $\vc{l}$ and $\vc{q}$)  the subset of the vectors associated with tetrahedron $i$.
Remaining quantities, $\mat{\vc{B}}\in\real^{3\times 3 \times 9}$, $\mat{\vc{C}}\in\real^{3\times 3 \times 6}$, $\mat{\vc{D}}^4\in\real^{3\times 4 \times 12}$, along with the tensor operations $\cdot$ and $:$, convert from vector variables to matrix variables via tensor contraction. Finally $\mat{G}^3_i\in\real^{4\times3}$ is the gradient operator for tetrahedron $i$.
See our supplemental material for their detailed definitions.

Extremizing the mixed finite element integration energy
\begin{equation}
    \vc{q}^{t+1}, \vc{s}^{t+1}, \vc{l} = \arg\max_\vc{l}\arg\min_{\vc{q},\vc{s}} \frac{1}{2h^2}\vc{a}\left(\vc{q}\right)^T\mat{M}\vc{a}\left(\vc{q}\right) + \tilde{U}^D\left(\vc{q},\vc{s},\vc{l}\right),
    \label{eq:integrate}
\end{equation} then gives us a forward time step update for a deforming mesh.

\subsection{A Newton-Like Solver}
Our fast dynamics solver~(Algorithm~\ref{alg:solver}) performs Newton-type iteration using a fast alternating projections solver to compute search directions.
The solver is applied to a quadratic approximation of~\autoref{eq:integrate} in $\vc{q}$ and $\vc{s}$, given by 

\begin{equation}
    \begin{split}
    \Delta E \rvert_{\vc{q}^k, \vc{s}^k} = & \frac{1}{2h^2}\Delta\vc{q}^T\mat{M}\Delta\vc{q} - \frac{1}{h^2}\Delta\vc{q}^TM\vc{a}\left(\vc{q}^k\right) \cdots \\
                                 & + \frac{1}{2}\Delta\vc{s}^T\mat{H}\Delta\vc{s} + \Delta\vc{s}^T\vc{g} \cdots \\
                                 & - \vc{l}^T\left(\mat{W}\left(\vc{s}^k+\Delta\vc{s}\right)-\mat{J}\left(\vc{q}^k+\Delta\vc{q}\right)\right).
    \end{split}
    \label{eq:incremental}
\end{equation} Here $\vc{q}^k$ and $\vc{s}^k$ are the current guesses for $\vc{q}$ and $\vc{s}$, $\mat{H}\in\real^{6|T|\times 6|T|}$ is a block diagonal matrix efficiently storing second-order elastic energy information, $\frac{\partial^2\psi}{\partial\vc{s}^2}\rvert_{\bar{\mat{P}}_i\vc{s}^k}$, in diagonal blocks, and $\vc{g}$ is the stacked vector of gradients, $\frac{\partial\psi}{\partial\vc{s}}\rvert_{\bar{\mat{P}_i}\vc{s}^k}$.
We compact our tensors from~\autoref{eq:mixed_discrete} into matrices; $\mat{J} = \sum_{i=1}^{|T|} \mat{P}_i^T\left(\mat{\vc{B}}^T:\left(G^{3}_i\mat{\vc{D}}^{4T}\right)\right)\mat{N}_i$, 
$\mat{W} = \sum_{i=1}^{|T|}\mat{\vc{Z}:\mat{R}_i}$ and $Z\in\real^{9\times6\times3\times3} = \mat{\vc{B}}\cdot\mat{\vc{C}}$.

\emph{Importantly} $\mat{J}$ is a constant matrix that can be assembled at initialization time and $\mat{W}$ is block diagonal which depends linearly on $\mat{R_i}$, and can be updated quickly, in parallel. 

Our method uses the search directions computed by minimizing \autoref{eq:incremental} in conjunction with a backtracking linesearch. 
Here the remaining critical ingredient is method to efficiently minimize \autoref{eq:incremental}.

\begin{algorithm}
    \caption{Our fast elastodynamics solver}\label{alg:solver}
    \begin{algorithmic}
    \Require $\vc{q}^t$, $\vc{q}^{t-1}$, $\vc{s}^t$
    \Require $n$ \Comment{maximum number of outer iterations}
    \Require $m$ \Comment{maximum number of inner iterations} 
    
    \State $ii \gets 0$
    \State $jj \gets 0$
    \State $\vc{l} \gets 0$
    \State $\vc{q^{t+1}} \gets 2\vc{q^t} - \vc{q^{t-1}} + h^2M^{-1}\vc{f}_{ext}$ \Comment{initalize with Forward Euler} 
    \State $\vc{s^{t+1}} \gets \vc{s^{t}}$
    
    \While{\mbox{Not Converged \textbf{and} } $ii<n$}
    
        \State \mbox{Update} $H$, $g$, $\vc{b}$
    
        \While{\mbox{Not Converged \textbf{and} }$jj$<$m$} \Comment{\S\ref{sec:alternating}}
            
            \State $\forall i \in |T|$, $\mat{R}_i \gets \mbox{procrustes}\left(\mat{N}_i\vc{q}^{t+1}, \bar{\mat{P}}_i\vc{s}^{t+1}, \mat{P}_i\vc{l}\right)$ \Comment{\S\ref{sec:local}}
    
            \State \mbox{Update} $\mat{W}$
    
            \State $\vc{q}^{t+1}, \vc{l} \gets \mbox{global}\left(\vc{q}^{t+1}, \vc{q}^{t}, \vc{q}^{t-1},\vc{s}^{t+1}\mat{R}_1 ... \mat{R}_{|T|}\right)$ \Comment{\S\ref{sec:global}}
            
            \State $\vc{s}^{t+1} \gets \mat{W}^{T}\vc{l}-\mat{H}^{-1}\vc{g}$ 
    
        \EndWhile
        
        \State $\vc{q}^{t+1}$, $\vc{s}^{t+1} \gets $ backtracking linesearch
    
    \EndWhile
    
    %\Result $\vc{q}^{t+1}$, $\vc{s}^{t+1}$
    
    \end{algorithmic}
\end{algorithm}

\subsection{Alternating Projections Solver}\label{sec:alternating}
While our formulation enables an efficient, sparse quadratic model in $\Delta\vc{q}$ and $\Delta\vc{s}$, it remains linear in our rotational variables. 
We apply alternating projections to minimize this per energy, motivated by prior successes when applied to similar objectives~\citep{Jacobson-12-FAST,Liu:2013:FSM,KKB2018}.
Each step of our alternating projections consists of a global substep updating $\Delta\vc{q}$ and $\Delta\vc{s}$ and a local substep updating per-element rotations. 

\subsubsection{Global Solve}\label{sec:global}

Optimality of \autoref{eq:incremental} with respect to $\Delta\vc{q}$ and $\Delta\vc{s}$ gives a standard symmetric indefinite KKT system~\citep{Wright:1999uv}, 
which we can simplify by eliminating $\vc{\Delta s}$. This recovers a generalized form of the compliant dynamic formulation of \citet{servin2006interactive}:

\begin{equation}
    \begin{pmatrix}
        \frac{1}{h^2}\mat{M} & \mat{J}^T \\
        \mat{J} & -\mat{W}\mat{H}^{-1}\mat{W}^T\\
    \end{pmatrix}
    \begin{pmatrix}
        \Delta\vc{q}\\
        \vc{l}
    \end{pmatrix}
    =
    \begin{pmatrix}
        \frac{1}{h^2}\mat{M}\vc{a} + \vc{f}_{ext} \\
        \mat{W}\vc{s}^k - \mat{J}\vc{q}^k + \mat{W}\mat{H}^{-1}\vc{g}
    \end{pmatrix}
    \label{eq:kkt}
\end{equation}

We apply preconditioned conjugate gradient to solve \autoref{eq:kkt}\ \citep{durazzi2003indefinitely}. 
Updating the left-hand side of the system only requires recomputing $-\mat{W}\mat{H}^{-1}\mat{W}^T$ which is block diagonal and can be done in parallel.
The right-hand side is similarly quick to update each iteration.
Our preconditioner is a constant form of \autoref{eq:kkt} with $\mat{W}^TH^{-1}\mat{W}=\frac{1}{\mu}I$ where $\mu$ is the first Lam\'{e} parameter of the simulated material.
This preconditioner is constant and so we prefactorize once for the entire simulation.
Comparable to fluid pressure solves, elastic finite element stresses have a nullspace. 
We find a small amount of Tikhonov Regularization ($10^{-6}$) added to the lower right-hand block of \autoref{eq:kkt} then resolves all related convergence issues. 
With $\vc{l}$ in hand we can compute $\Delta\vc{s} = \mat{H}^{-1}\left(\mat{W}^T\vc{l} - \vc{g}\right)$ in parallel for each tetrahedron.

\subsubsection{Local Rotation Update}\label{sec:local}

Minimizing \autoref{eq:incremental} directly for rotation variables leads to an unstable algorithm so instead our local rotation update 
minimizes the augmented Lagrangian form with quadratic penalty. 
This yields a fast, stabilized rotation update which is computed by solving
\begin{equation}
    \mat{R}_i = \arg\max_{\mat{R}} \left<\mat{R}, \left(\frac{1}{\beta}\lambda_i + \mat{F}_i\right)\mat{S_i}^T\right>_F,
\end{equation} where $\left<\cdot,\cdot\right>_F$ is the Frobenius inner product, $\lambda_i=\mat{\vc{B}}\cdot\left(\mat{P}_i\vc{l}\right)$, $\mat{F}_i = \mat{\vc{B}}\cdot\left(\mat{P}_i\mat{J}\left(\vc{q^k}+\vc{\Delta\vc{q}}\right)\right)$ and 
$\mat{S}_i = \mat{\vc{C}}\cdot\bar{\mat{P}}_i\left(\vc{s}^k+\Delta s\right)$.
This orthogonal Procrustes problem~\citep{eggert1997estimating} can be solved efficiently via fast Singular Value Decomposition~\citep{10.1145/1964921.1964932}, parallelized over each tetrahedron.
Here $\beta$ is the quadratic penalty parameter. 
Following guidance from~\citet{Wright:1999uv} that $\beta$ should be greater than $|\mat{P}_i\vc{l}|_{\infty}$ to guarantee convergence, we set $\beta = \alpha\cdot|\mat{P}_i\vc{l}|$, where $\alpha = 10$ initially and increases by $1.5\times$ each substep.

\subsection{Order of Local-Global Solves and Initial Guess}
Our choice of ordering and initial guess is taken from previous work. 
Specifically both \citet{10.1145/1073204.1073216} and \citet{Li:2019:DOT} advise warm-starting backwards Euler via the forward Euler predictor while Shape Matching~\citep{10.1145/1073204.1073216} finds an initial, best-fit rigid rotation to this initial guess. Forward Euler initialization, combined with performing the local rotation first in the alternating projections sequence, effectively generalizes the Shape Matching predictor-corrector approach.

\subsection{Constraints}
Pinned vertices are handled by standard projection from global position vectors, $\vc{q}$, in all solves. 
For contact in our current proof-of-concept implementation we apply penalty springs~\citep{10.1145/566654.566623} with hand tuned parameters for each example.
We do not handle self-collisions in the current implementation.

\subsection{Cloth and Rods}
Our mixed formulation is trivially adaptable to cloth and rod simulation, requiring only four minor updates to the above algorithm:
\begin{enumerate}
    \item Rebuild $\mat{J}$ using $\mat{\vc{D}}^3$ (resp. $\mat{\vc{D}}^2$) and $\mat{\vc{G}}^2$ (resp. $\mat{\vc{G}}^1$) for cloth (rods).
    \item For cloth (resp. rods), add $\sum_{i=1}^{i=|T|}\mat{P}_i^T\mat{\vc{B}}:(\mat{R}_i\vc{n}_i\vc{n}_i^T)$ \\ (additionally $\sum_{i=1}^{i=|T|}\mat{P}_i^T\mat{\vc{B}}:(\mat{R}_i\vc{n}'_i\vc{n}^{'T}_i)$ for rods) to the Lagrange multiplier right-hand side in \autoref{eq:kkt}. Here $\vc{n}_i$ is the $i^{th}$, per-facet reference space normal, and $\vc{n}'_i$ the binormal.
    \item Replace $\mat{S}_i$ with $\mat{S}_i - \vc{n}_i\vc{n}_i^T$ (resp. $\mat{S}_i - \vc{n}\vc{n}^T - \vc{n'}\vc{n}^{'T}$) in the local step for cloth (rods).
    \item Replace integration volumes~(\autoref{eq:mixed_discrete}) with appropriate values.
\end{enumerate}

\emph{Crucially} we do not need to alter the strain energy density calculation in any way when simulating rods or cloth.
No matter what the input discretization, material properties are specified using standard, volumetric models, which avoids complications when moving between different representations. 
This also enables interesting generalizations (ie fast simulation of nonlinear, Neohookean springs).

\section{Results and Discussion}
All our experiments are performed on a MacBook Pro 13.3" (Apple M1 8-core CPU, 8-core GPU, 16GB Memory, 512GB SSD). We implement our method with Eigen~\citep{eigenweb} for linear algebra routines, SuiteSparse for direct linear solves~\citep{davis2006direct}, libigl~\citep{libigl} for geometry processing, Bartels~\citep{bartels} for physics utility code and Polyscope~\cite{polyscope} for display.
Due to time constraints optimizations were limited to obvious precomputation and multithreading opportunities.
For all examples we use an interactive configuration with a fixed number of outer and inner iterations (Algorithm \autoref{alg:solver}, $n=1$, $m$ given in \autoref{tbl:perf}).
We use a  residual tolerance of $1e-7$ for our Conjugate Gradient solver, which is currently not assembly free.
Despite the relatively low iteration counts our method generates a myriad of visually plausible results quickly~(\autoref{tbl:perf}) and robustly. 

  \begin{table*}[h]
  \setlength{\tabcolsep}{3pt}
  \caption{Simulation statistics for all examples. Material parameters of density ($\rho$), Young's Modulus, (\textbf{E}), Poisson's ratio ($\nu$), and material model(\textbf{Model}) options, neo-Hookean (NH), corotational (Corot), or ARAP, specified per example.  We report wall-clock timings in milliseconds (ms) with   \textbf{Substeps} the number of global substeps  per timestep. Here, the provided timings give the time taken in a single substep.
   \textbf{Assembly} is the time to assemble the left-hand side KKT system and the corresponding right-hand side;
   \textbf{KKT Solve} is the time to solve the KKT system; and \textbf{Rotation Solve} is the time taken for the local rotation update plus the time to compute and apply $\Delta\vc{s}$.}
  \label{tbl:perf}
  \begin{center}
  \begin{tabular}{l c c c c c c c c c c c}
   \textbf{Example} & $|\vc{q}|$ & |T| &  \textbf{Model} & $\rho$(kg/m$^3$) & \textbf{E}(Pa) & $\nu$ &
	\textbf{Substeps} & \textbf{Assembly (ms)} & \textbf{KKT Solve (ms)} & \textbf{Rotation Solve (ms)} \\
   \hline 
   \rowcolor[HTML]{DAE8FC}
   \textbf{Square Cloth (soft)}  & 5929 & 11552 & NH & $1e^2$ & $1e^5$& 0.40 & 5 & 5.40 & 3.69 & 1.95  \\
   \textbf{Square Cloth (stiff)} & 5929 & 11552 & NH & $1e^3$ & $1e^9$& 0.40 & 5 & 3.47 & 3.37 &  2.03 \\
   \rowcolor[HTML]{DAE8FC}
  \textbf{Rod Armadillo} & 675 & 2019 & NH & $1e^1$ & $1e^7$& 0.45 & 10 & 1.01 & 0.48 &  0.73 \\
  \textbf{Jelly Octopus} & 452 & 1140 & Corot & $1e^3$ & $5e^5$& 0.45 & 3 & 0.36 & 0.71 &  0.29 \\
     \rowcolor[HTML]{DAE8FC}
  \textbf{Beam ARAP} & 6000 & 29142 & ARAP & $1e^3$ & $1e^5$& 0.45 & 5 & 8.24 & 14.9 &  3.06\\
  \textbf{Beam Corot} & 6000 & 29142 & Corot & $1e^3$ & $1e^5$& 0.45 & 5 & 4.74 & 82.6 &  3.24 \\
     \rowcolor[HTML]{DAE8FC}

  \textbf{Bean NH} & 6000 & 29142 & NH & $1e^3$ & $1e^5$& 0.45 & 5 & 8.88 & 169 &  3.21 \\
  \textbf{Tet Bunny} & 699 & 2274 & NH & $1e^3$ & $1e^5$& 0.45 & 5 & 0.90 & 1.47 &  0.33 \\
     \rowcolor[HTML]{DAE8FC}

  \textbf{Cloth Bunny} & 34834 & 69664 & NH & $1e^2$ & $1e^5$& 0.40 & 5 & 17.38 & 16.3 &  7.18 \\
  \textbf{Rods Bunny} & 500 & 2434 & NH & $1e^2$ & $2e^5$& 0.45 & 5 & 0.95 & 1.19 &  0.44 \\
     \rowcolor[HTML]{DAE8FC}

  \textbf{Spot Drop (soft)} & 4079 & 15555 & Corot & $1e^3$ & $1e^6$& 0.45 & 15 & 2.29  & 25.0 &  1.80 \\
       %\rowcolor[HTML]{DAE8FC}

  \textbf{Spot Drop (stiff)} & 4079 & 15555 & Corot & $1e^3$ & $1e^9$& 0.45 & 15 & 1.28 & 4.02 &  1.83 \\
   \hline
  \end{tabular}
  \end{center}
  \end{table*}

\begin{figure*}[htp]
    \includegraphics[width=.93\textwidth]{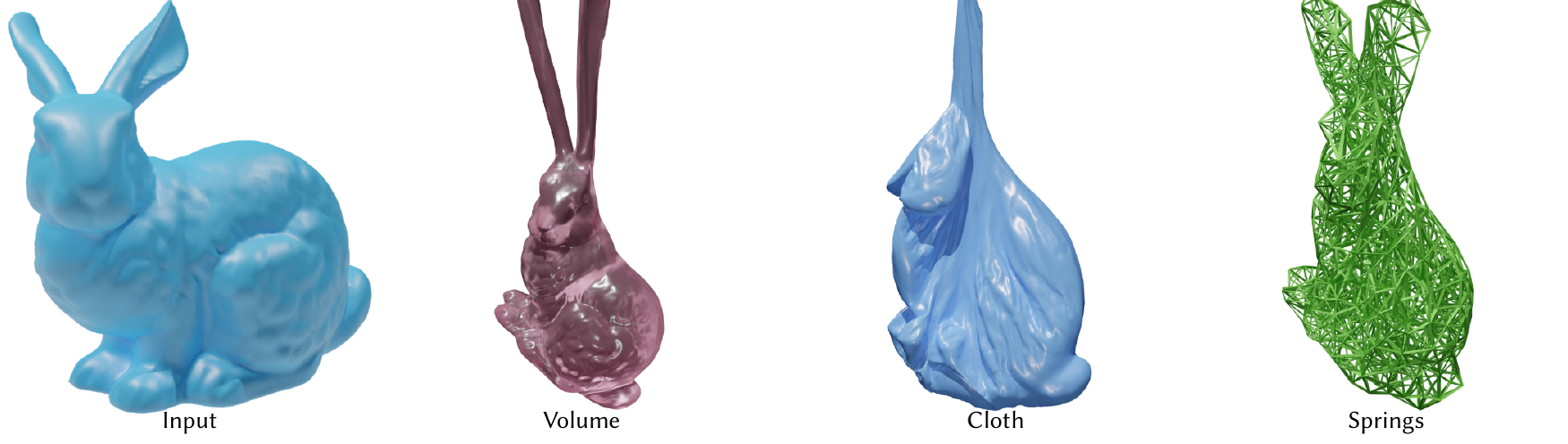}
    \caption{Stanford Bunny Three Ways: The Stanford Bunny simulated as a volume, a shell and springs using the Neohookean material model.}
    \label{fig:reps}
  \end{figure*}

  Our solver is cable of simulating volumetric objects, surface only objects and objects made of rods/springs~(\autoref{fig:reps}). 
  We show the same Stanford Bunny mesh simulated using tetrahedral finite elements, triangular finite elements for surface simulation, and springs placed along the edges of a coarse tetrahedral mesh.
  The volumetric FEM simulation has been skinned with a high resolution surface mesh. 
  We use a neo-Hookean material model for all three simulations.

\begin{figure}[h]
    \includegraphics[width=.93\columnwidth]{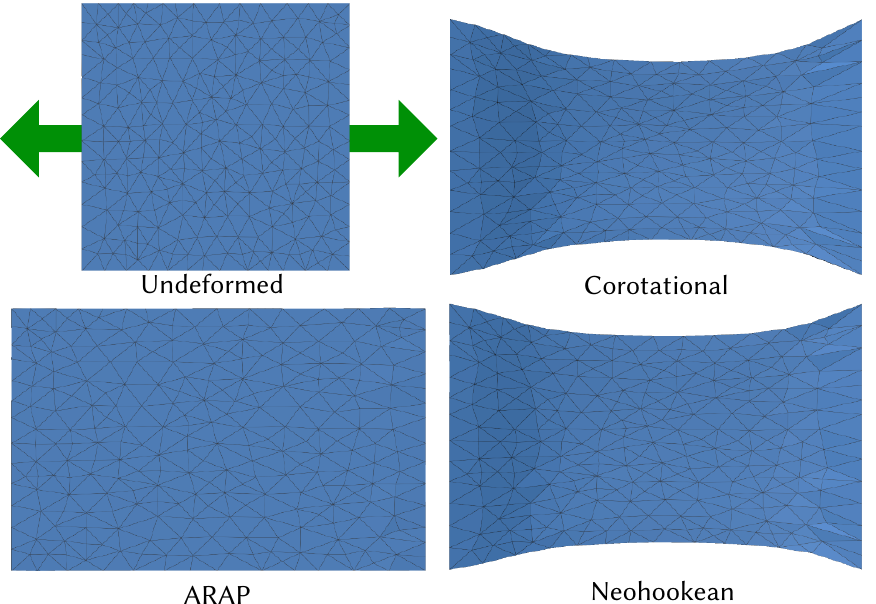}
    \caption{Simulating stretching a square using ARAP, corotational and neo-Hookean hyperelastic models.}
    \label{fig:materials}
  \end{figure}

  We apply three different materials (ARAP, Corotational elasticity and Neohookean elasticity) to identical squares, which are then stretched via moving symmetric boundary conditions on the left and right sides~(\autoref{fig:materials}).
  Because the ARAP model omits a volume preservation term, its deformation is purely orthogonal while corotational and neo-Hookean material models exhibit varying degrees of necking.

\begin{figure}[h]
    \includegraphics[width=.93\columnwidth]{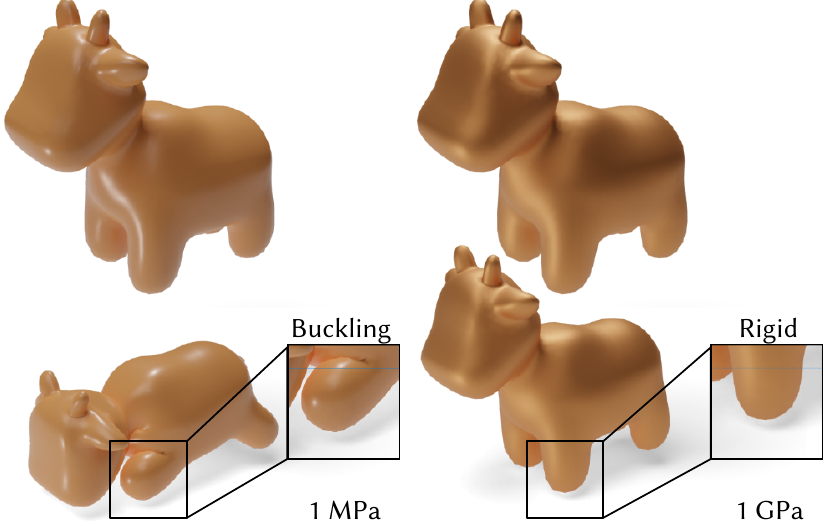}
    \caption{Spot Drop Test: We simulate two cows being dropped from identical heights onto a ground plane. Our method is equally effective for soft ($1$MPa) and  stiff ($1$GPa) materials}.
    \label{fig:compliance}
  \end{figure}

  Next we demonstrate our solver's robustness in the face of changing material properties~(\autoref{fig:compliance}). 
  Here two cows are both simulated using the neo-Hookean material model but with vastly different stiffnesses($1$GPa vs $1$MPa).
  In both cases we observe stable, plausible simulations with expected differences in deformation behavior -- importantly the stiff cow remains close to rigid, even with our frugal iteration limit.

\section{Conclusion and Future Work}
We have demonstrated how a strategically designed mixed variational potential and corresponding finite element discretization lead us to an efficient, robust, and flexible solver for elastodynamics.
In order to promote further exploration of our approach we will release an open source version of our solver, under a permissive license.
This new simulation framework, as shown above, already offers exciting simulation possibilities, while likewise elucidating connections to both popular  frameworks like PBD, as well as to more traditional finite element approaches. 
While the method derived here has compelling features and performance, we see this as first steps in exploring opportunities for mixed variational formulations. 
One immediate and attractive opportunity is to replace our preconditioned conjugate gradient solver with a customized multigrid approach. This could amount to applying a multi-resolution dynamics solver for the global step of our method.
Integrating more robust contact handling, possibly via recent incremental potential contact~\cite{Li2020IPC} approaches, will also go a long way towards broadening the practical application of the current method and could also open the door to more extended mixed-field models.
Finally, further exploration of the connections between our mixed-variational approach and alternate approaches to splitting in numerical optimization such as Douglas-Rachford, should be a promising avenue to explore for further improved convergence and performance of fast deformable-body simulation methods.

\bibliographystyle{ACM-Reference-Format}
\bibliography{references}

%\appendix 
%\input{sections/appendix}

\end{document}

\title{Supplemental Material for Mixed Variational Finite Elements for Implicit, Position-Based Simulation of Solid Bodies}

\author{Ty Trusty}
\affiliation{%
  \institution{University of Toronto}
  \country{Canada}
}

\author{Danny M. Kaufman}
\affiliation{%
  \institution{Adobe Research}
  \country{USA}
}

\author{David I.W Levin}
%\authornotemark[3]
\affiliation{%
  \institution{University of Toronto}
  \country{Canada}
}

\renewcommand{\shortauthors}{Levin}

\maketitle

\section{List of Tensors}\label{sec:tensors}

\[
\mat{\vc{B}}\in\real^{3\times3\times9}= \mat{\vc{B}}_{ijk} = 
\begin{bmatrix}
    %\begin{rcases}
        \begin{pmatrix}
            1 & 0 & 0 & 0 & 0 & 0 & 0 & 0 & 0\\
            0 & 1 & 0 & 0 & 0 & 0 & 0 & 0 & 0\\
            0 & 0 & 1 & 0 & 0 & 0 & 0 & 0 & 0\\
        \end{pmatrix}
    i=0,jk\\
    \begin{pmatrix}
        0 & 0 & 0 & 1 & 0 & 0 & 0 & 0 & 0\\
        0 & 0 & 0 & 0 & 1 & 0 & 0 & 0 & 0\\
        0 & 0 & 0 & 0 & 0 & 1 & 0 & 0 & 0\\
    \end{pmatrix}i=1,jk\\
    \begin{pmatrix}
        0 & 0 & 0 & 0 & 0 & 0 & 1 & 0 & 0\\
        0 & 0 & 0 & 0 & 0 & 0 & 0 & 1 & 0\\
        0 & 0 & 0 & 0 & 0 & 0 & 0 & 0 & 1\\
    \end{pmatrix}i=2,jk\\
\end{bmatrix}
\]
\[
\mat{\vc{C}}\in\real^{3\times3\times6} = \mat{\vc{C}}_{ijk} = 
\begin{bmatrix}
    %\begin{rcases}
        \begin{pmatrix}
            1 & 0 & 0 & 0 & 0 & 0\\
            0 & 0 & 0 & 0 & 0 & 1\\
            0 & 0 & 0 & 0 & 1 & 0\\
        \end{pmatrix}
    i=0,jk\\
    \begin{pmatrix}
        0 & 0 & 0 & 0 & 0 & 1\\
        0 & 1 & 0 & 0 & 0 & 0\\
        0 & 0 & 0 & 1 & 0 & 0\\
    \end{pmatrix}i=1,jk\\
    \begin{pmatrix}
        0 & 0 & 0 & 0 & 1 & 0\\
        0 & 0 & 0 & 1 & 0 & 0\\
        0 & 0 & 1 & 0 & 0 & 0\\
    \end{pmatrix}i=2,jk\\
\end{bmatrix}
\]
\[
\mat{\vc{D}}^n\in\real^{3\times n\times 3*n} = \mat{\vc{D}}_{ijk} = \delta_{i,mod(k,3)}\delta_{3j,k-i},
\] where $\delta_{ij}$ is the Kroenecker Delta.

\textbf{Example}
\[
\mat{\vc{D}}^3\in\real^{3\times3\times9} = \mat{\vc{D}}_{ijk} =
\begin{bmatrix}
    %\begin{rcases}
        \begin{pmatrix}
            1 & 0 & 0 & 0 & 0 & 0 & 0 & 0 & 0\\
            0 & 0 & 0 & 1 & 0 & 0 & 0 & 0 & 0\\
            0 & 0 & 0 & 0 & 0 & 0 & 1 & 0 & 0\\
        \end{pmatrix} i=0,jk\\
    \begin{pmatrix}
        0 & 1 & 0 & 0 & 0 & 0 & 0 & 0 & 0\\
        0 & 0 & 0 & 0 & 1 & 0 & 0 & 0 & 0\\
        0 & 0 & 0 & 0 & 0 & 0 & 0 & 1 & 0\\
    \end{pmatrix} i=1,jk\\
    \begin{pmatrix}
        0 & 0 & 1 & 0 & 0 & 0 & 0 & 0 & 0 \\
        0 & 0 & 0 & 0 & 0 & 1 & 0 & 0 & 0 \\
        0 & 0 & 0 & 0 & 0 & 0 & 0 & 0 & 1 \\
    \end{pmatrix} i=2,jk\\
\end{bmatrix}
\]
\vspace{3ex}
\section{List of Tensor Operations}\label{sec:operations}
\[
    \mat{A}\mat{\vc{B}} = \mat{\vc{C}}_{ijk} = \sum_{l=1}^{\mbox{col}\left(\mat{\vc{A}}\right)} \mat{A}_{il}\mat{\vc{B}}_{ljk}
\] 

\[
    \mat{\vc{A}}^T = \mat{\vc{C}}_{ijk} = \mat{\vc{A}}_{jik}
\] 
\[
     \mat{\vc{A}}:\mat{\vc{B}} = \mat{C}_{kl} = \sum_{i=1}^{\mbox{rows}\left(\mat{\vc{A}}\right)} \sum_{j=1}^{\mbox{cols}\left(\mat{\vc{A}}\right)}\mat{A}_{ijk}\mat{B}_{ijl}
\]
\[
     \mat{\vc{A}}:\mat{B} = \vc{c}_{k} = \sum_{i=1}^{\mbox{rows}\left(\mat{\vc{A}}\right)} \sum_{j=1}^{\mbox{cols}\left(\mat{\vc{A}}\right)}\mat{A}_{ijk}\mat{B}_{ij}
\] 
\[
    \mat{\vc{A}}\cdot\vc{b} = (\mat{\vc{A}}\cdot\vc{b})_{ij} = \sum_{k=1}^{\mbox{depth}\left(\mat{\vc{A}}\right)} \mat{\vc{A}}_{ijk}\cdot\vc{b}_k
\]
\[
    \mat{\vc{A}}\cdot\mat{\vc{B}} = \mat{\vc{C}}_{ijkl} =  \sum_{m=1}^{\mbox{cols}\left(\mat{A}\right)}\mat{\vc{A}}_{kmi}\mat{\vc{B}}_{lmj}
\]
\[
    \mat{A}:\mat{B} = \sum_{i=1}^{\mbox{rows}\left(\mat{A}\right)} \sum_{j=1}^{\mbox{cols}\left(\mat{A}\right)}\mat{A}_{ij}\mat{B}_{ij}
\]

\section{Gradients}\label{sec:gradients}

\textbf{Tetrahedron Gradient}
\[
    \mat{G}^3 \in\real^{4\times 3} = 
    \begin{pmatrix}
        -\mathbbm{1}\mat{T}^{-1} \\
        \mat{T}^{-1}
    \end{pmatrix},
\] where $T\in\real^{3\times3} = \left(\Delta\vc{X}_0, \Delta\vc{X}_1, \Delta\vc{X}_2\right)$, $\Delta\vc{X}_{0,1,2}$ are the edge vectors for a tetrahedron and $\mathbbm{1}=\left(1,1,1\right)$.

\textbf{Triangle Gradient}
\[
    \mat{G}^2 \in\real^{3\times 3} = 
    \begin{pmatrix}
        -\mathbbm{1}\left(\mat{T}^{T}\mat{T}\right)^{-1}\mat{T^{T}} \\
        \left(\mat{T}^{T}\mat{T}\right)^{-1}\mat{T^{T}}
    \end{pmatrix},
\] where $T\in\real^{2\times3} = \left(\Delta\vc{X}_0,\Delta\vc{X}_1\right)$, $\Delta\vc{X}_{0,1}$ are the edge vectors for a triangle and $\mathbbm{1}=\left(1,1\right)$.

\textbf{Piecewise Linear Curve Gradient}

\[
    \mat{G}^1 \in\real^{2\times 3} = 
    \begin{pmatrix}
        -\left(\mat{T}^{T}\mat{T}\right)^{-1}\mat{T^{T}} \\
        \left(\mat{T}^{T}\mat{T}\right)^{-1}\mat{T^{T}}
    \end{pmatrix},
\] where $T\in\real^{1\times3} = \Delta\vc{X}$ is the edge vector for a rod.